\documentstyle[12pt,epsf]{article}    
\newcommand{\comment}[1]{}
\newcommand{\PSbox}[3]{\mbox{\rule{0in}{#3}
\includegraphics{#1}\hspace{#2}}}

\newcommand{\refb}[1]{(\ref{#1})}

\newcommand{\wh}{\widehat}

\def\bbbz{{\sf Z\!\!\!Z}}

\def\sl2z{SL(2,\bbbz)}
\newcommand{\be}{\begin{equation}}
\newcommand{\ee}{\end{equation}}
\newcommand{\bea}{\begin{eqnarray}}
\newcommand{\eea}{\end{eqnarray}}
\newcommand{\nn}{\nonumber}

\newcommand{\hp}{{\wh\Phi}}
\newcommand{\hq}{{\wh Q_B}}
\newcommand{\he}{{\wh\eta_0}}
\newcommand{\ha}{{\wh{A}}}

\newcommand{\lllb}{\Bigl\langle\Bigl\langle}
\newcommand{\rrrb}{\Bigr\rangle\Bigr\rangle}

\newcommand{\ep}{e^{-\phi}}
\newcommand{\de}{\partial}

\def\Tr{\rm Tr\ }

\def\bbbz{{\sf Z\!\!\!Z}}
\def\sl2z{SL(2,\bbbz)}
\def\fracs#1#2{\textstyle\frac #1#2}

\def\z0{{\bf z_0}}

\newcommand{\rrr}{\rangle \rangle}
\newcommand{\lll}{\langle \langle}

\newcommand{\p}{\partial }
\begin{document}
\newpage

\noindent
\thispagestyle{empty}
{\flushright{\small MIT-CTP-2968\\hep-th/0004015\\}}

\vspace{.3in}
\begin{center}\Large {\bf Tachyon Condensation on a Non-BPS D-Brane}

\end{center}

\vspace{.3in}
\begin{center}
{\large Amer Iqbal  \,and\,  Asad Naqvi}
\vspace{.1in}

Center for Theoretical Physics,\\
Massachusetts Institute of Technology,\\
Cambridge, Massachusetts 02139, U.S.A.
\vspace{.2in}

E-mail: {\tt iqbal@ctpbronze.mit.edu, asad@ctppurple.mit.edu}
\end{center}

\vspace{0.1in}

\begin{abstract}
We extend the recent computation of the tachyon potential by Berkovits, Sen 
and Zwiebach
by including level two fields and keeping up to level four terms in the
action. We find 90.5\% 
of the expected result. 
\end{abstract}
\newpage

\section{Introduction}
During the last few years, it has been realized 
\cite{BG,AS1,AS2,AS3} that 
Type IIA (IIB) string theory contains non-BPS D-branes
of odd (even) spatial dimensions. An analysis of the
open string states living on such a brane reveals the
existence of a tachyon which comes from the NS sector. 
It has been argued by Sen \cite{9805019,9805170} that the tachyon
potential has a minimum and that the minimum corresponds
to the usual vacuum of closed string theory without any
D-brane. This implies that the negative energy density
contribution from the tachyon potential must exactly cancel
the tension of the non-BPS D-brane. 

To analyze the tachyon potential in order to test this 
conjecture, ordinary first quantized
string theory cannot be used since the zero momentum tachyon
state is not on-shell. We need an off-shell formalism
of string theory. Such a formalism is given by string field theory. 
Recently it has been realized that calculating the tachyon potential
in string field theory by including fields and terms in the
string field action up to a certain level gives a good 
approximation to the tachyon potential.
In 
\cite{0001084,bsz}, the zeroth order and the first non-trivial
correction to the tachyon potential on a non-BPS D-brane of
Type II string theory was computed using open string field
theory action formulated in \cite{9503099,9912121}. It was found
that the potential has a minimum which cancels 60\%
of the D-brane tension at the zeroth level and 85\%
of the D-brane tension at the next non-trivial level (level 3/2). In this
paper, we extend the work of \cite{bsz} to include fields
up to level 2. Similar computation was done in \cite{DS}.   

This paper is organized as follows: In section 2, we review the
formalism of \cite{0001084,bsz} to set up the calculation. In section 3, 
we write down the expansion of the string field including fields
up to level 2. We also give some details of the 
calculation --- we give one simple example of calculation of a term in the 
action at each level.  In the appendix, we list the conformal and BRST
transformations of operators that we need in the calculation.  

\section{Superstring field theory on a Non-BPS D-brane}
Superstring field theory on a non-BPS D-brane was described in detail
in \cite{bsz}. Here, we will just review the basic features to
set up the calculation and establish notation. We will first review
the essential features of string field theory on a BPS D-brane which
only has a GSO(+) sector and then introduce Chan Paton matrices and
GSO($-$) sector to discuss the non-BPS D-brane. We follow the notation
and conventions of \cite{bsz} throughout the paper. 

\subsubsection*{BPS D-brane}
In \cite{9503099,9912121,0001084}, a  string field
configuration in the GSO$(+)$ NS sector corresponds to 
a Grassmann even open string vertex operator
$\Phi$ of ghost and picture number 0 \cite{FMS} in the combined
conformal field theory of a $c=15$ superconformal matter system, and
the $b,c,\beta,\gamma$ ghost system with $c=-15$. Here, $b$, $c$ are
the reparameterization ghosts and $\beta$, $\gamma$ are their superconformal
partners. The $\beta$, $\gamma$ ghost system can be bosonized and can 
be replaced by ghost fields $\xi,\eta,\phi$, related to $\beta,\gamma$ through
the relations 
\[ 
\beta=\de\xi e^{-\phi}, \qquad \gamma =\eta e^\phi.
\]
The ghost number ($n_g$) and picture number ($n_p$) assignments and
the conformal weights ($h$) are as
follows \cite{bsz}:
\begin{eqnarray*}
&& b:\quad n_g=-1, n_p=0,h=2 \qquad c:\quad n_g=1, n_p=0,h=-1 \, , \nonumber \\
&& e^{q\phi}:\quad n_g=0, n_p=q,h=-(q+\frac{q^2}{2})\, , \nonumber \\
&& \xi:\quad n_g=-1, n_p=1,h=0, \qquad \eta:\quad n_g=1, n_p=-1,h=1\, . \nonumber
\end{eqnarray*} 
The SL(2,R) invariant vacuum has zero ghost and picture number.

We shall denote by $\langle \prod_i A_i \rangle$ the correlation function
of a set of vertex operators in the combined matter-ghost conformal field
theory on the unit disk
with open string vertex operators inserted on the boundary of the disk,
without including trace over CP factors.
These
correlation functions are to be computed with the
normalization
\be \label{eb1}
\langle\xi(z) c\p c\p^2 c(w) e^{-2\phi(y)}\rangle = 2\, .
\ee 
The BRST operator is given by 
\[
Q_B=Q_0 + Q_1+Q_2,
\]
where
\bea 
\label{BRST}
Q_0& =&
= \oint dz   c \bigl( T_m + T_{\xi\eta} + T_\phi)
+ c \partial c b \nn, \\
Q_1& = &  \oint dz \eta \,e^\phi
\, G_m,  \,\,\,\,\ Q_2  =  -\oint dz \eta\partial \eta e^{2\phi} b \,.
\eea
Here
\[ 
\label{xietaphi}
T_{\xi\eta}=\partial\xi\,\eta, \quad T_\phi=-\fracs{{1}}{{2}} \partial\phi \partial
\phi -\partial^{2}\phi \, ,
\]
$T_m$ is the matter stress tensor and 
$G_m$ is the matter
superconformal generator. $G_m$ is a dimension $3/2$ primary field and 
satisfies:
\[ 
G_m(z) G_m(w) \simeq  {10\over (z-w)^3} + {2T_m\over (z-w)}\, .
\]
The
normalization of
$\phi$, $\xi$, $\eta$, $b$ and $c$ are as follows:
\[
\xi(z) \eta(w) \simeq {1 \over z-w}, \quad b(z) c(w)\simeq {1\over z-w}, 
\quad \partial \phi(z)
\partial\phi(w)\simeq
-{1\over (z-w)^2} \, .
\]
We denote by $\eta_0=\oint dz \eta(z)$ the zero mode of 
the field $\eta$ acting on the Hilbert space of matter ghost CFT.

The string field theory action  is
given by\cite{0001084}
\be \label{e0}
S={1\over 2g^2} \lllb (e^{-\Phi} Q_B e^{\Phi}) 
(e^{-\Phi}\eta_0 e^\Phi)
- \int_0^1 dt 
(e^{-t\Phi}\p_t e^{t\Phi})\{ (e^{-t\Phi}Q_B e^{t\Phi}),
(e^{-t\Phi}\eta_0 e^{t\Phi})\}\rrrb\,. 
\ee 
 This action is defined  
by expanding all exponentials in a formal Taylor series.
$\lll A_1, \ldots A_n \rrr$
is defined as:
\be \label{e2ff}
\lll A_1\ldots A_n \rrr = \Bigl\langle f^{(n)}_1 \circ A_1(0)\cdots 
f^{(n)}_n\circ A_n(0)\Bigr\rangle\, .
\ee
Here, 
$f\circ A$ for
any function $f(z)$, denotes the conformal transform of $A$ by $f$, and 
\be \label{e3}
f^{(n)}_k(z) = e^{2\pi i (k-1)\over n} \Big({1+iz\over 1-iz}
\Big)^{2/n}\,\quad  \hbox{for} \quad n\geq 1 .   
\ee
 In
particular if
$\varphi$
denotes a primary field of weight
$h$, then
\[
f\circ \varphi(0) = (f'(0))^h \varphi(f(0))\, .
\] 
Since we have, in general, non-integer weight vertex operators, we
should be more careful in defining $f\circ A$ for such vertex
operators.
Noting that 
\[ 
f^{(N)\prime}_k(0) = {4 i\over N} e^{2 \pi i{k-1 \over N}} \equiv {4\over
N}
e^{2\pi i ({k-1\over N} + {1\over 4})}\, ,
\]
we adopt the following definition of $f^{(N)}_k\circ\varphi(0)$ for a
primary
vertex operator $\varphi(x)$ of conformal weight $h$:
\be \label{e3d}
f^{(N)}_k\circ\varphi(0) = \bigg|\Big({4\over N}\Big)^h\bigg| e^{2\pi i h
({k-1 \over N}+{1\over
4})}\varphi(f^{(N)}_k(0))\, .
\ee

We also recall
identities \cite{bsz} which will be used in the calculation later:
\[
\{ Q_B, \eta_0\} = 0, \quad Q_B^2 = \eta_0^2 = 0 ,
\]
\be \label{epr1}
Q_B (\Phi_1 \Phi_2) = (Q_B\Phi_1)\Phi_2 + 
\Phi_1 (Q_B\Phi_2),\quad
\eta_0 (\Phi_1\Phi_2) = 
(\eta_0\Phi_1)\Phi_2 + \Phi_1 (\eta_0\Phi_2), 
\ee
\[
\lll Q_B (...) \rrr =
\lll \eta_0(...) \rrr =0.
\]
Note that in the identities of the
second line there are no minus signs necessary as $Q_B$ or 
$\eta_0$ go through the string field because the string
field is Grassmann even (since we are in the GSO($+$) 
sector only).

\subsubsection*{Non-BPS D-brane}
The algebraic structure described in the previous section works for
the GSO($+$) (Grassman even) sector living on a BPS D-brane. However, on a non-BPS
D-brane, the open string states live in both the GSO($-$) and GSO($+$)
sector. The GSO($-$) states are Grassman odd and to incorporate them 
in the algebraic structure of the previous section, internal $2 \times 2$
Chan Paton matrices were introduced in \cite{bsz}. These are added
both to the vertex operators and $Q_B$ and $\eta_0$ as described in 
detail below.

The  $2\times 2$ identity matrix $I$ is attached  
on the usual GSO$(+)$ sector  and  the
Pauli matrix $\sigma_1$ to the  GSO$(-)$ sector. The complete
string field is thus written as
\[
\hp = \Phi_+ \otimes I  + \Phi_- \otimes \sigma_1\,,
\]
where the subscripts denote the $(-)^F$ eigenvalue of the
vertex operator.   In addition, $Q_B$ and $ \eta_0$ are tensored 
with $\sigma_3$:
\[ 
\wh Q_B = Q_B\otimes \sigma_3, \qquad
\wh \eta_0 = \eta_0\otimes \sigma_3\, .
\]
We define
\be \label{e200}
\lll \wh A_1\ldots \wh A_n \rrr = {\Tr} \Bigl\langle f^{(n)}_1 \circ
\ha_1(0)\cdots
f^{(n)}_n\circ \ha_n(0)\Bigr\rangle \, ,
\ee
where the trace is over the internal Chan Paton matrices.
As in \cite{bsz}, fields or operators with internal
CP matrices
included are denoted by symbols with a
hat on them, and fields or operators without internal
CP matrices included are
denoted by symbols without a hat. 

In addition, we have the analogs of \refb{epr1} holding:
\[
\{\hq ,\he\} = 0, \quad  \hq^2 = \he^2 =0,
\]
\be \label{epr2a} 
\hq (\wh\Phi_1 \wh\Phi_2) = (\hq\wh\Phi_1)\wh\Phi_2 + 
\wh\Phi_1 (\hq\wh\Phi_2),\quad
\he (\wh\Phi_1 \wh\Phi_2) = 
(\he\wh\Phi_1)\wh\Phi_2 + \wh\Phi_1 (\he\wh\Phi_2), 
\ee
\[
\lll \hq(...) \rrr =
\lll \he(...) \rrr =0.
\]
The reason no extra signs appear in the middle equation is that
when the string field is Grassmann odd  
the sign arising by moving $Q_B$ across
the vertex operator is canceled by 
having to move $\sigma_3$ across $\sigma_1$.

Given that the relations satisfied by the hatted objects
are the same as those of the unhatted ones, 
the string field  action for the non-BPS D-brane takes the  
same structural form as that in \refb{e0} and is given by \cite{0001084}:
\be \label{e00}
S={1\over 4g^2} \lllb (e^{-\hp} \hq e^{\hp})(e^{-\hp}\he e^\hp) -
\int_0^1 dt (e^{-t\hp}\p_t e^{t\hp})\{ (e^{-t\hp}\hq e^{t\hp}),
(e^{-t\hp}\he e^{t\hp})\}\rrrb\, ,  
\ee 
Here the overall normalization is divided  by a factor
of two in order to compensate for the trace operation on the
internal matrices.

\section{Tachyon Potential and Level Approximation}  
To study the phenomenon of tachyon condensation on the non-BPS
D-brane,
 we will restrict our string field to be in 
${\cal H}_1$ \cite{bsz, sen}, the subset of vertex operators of ghost and 
picture number zero, created from the matter stress tensor ($T_m(z)$), 
its superconformal partner ($G_m(z)$) and the ghost fields $b,c,\xi, \eta,
\phi$.  The level of a string field 
component multiplying a vertex operator of conformal weight $h$ is defined to be 
$(h+\fracs{{1}}{{2}})$, so that the tachyon field multiplying the
vertex operator $\xi ce^{-\phi}\otimes\sigma_1$, has level zero. The
level of a given term in the string field  action is defined to be the sum of the levels
of the individual fields appearing in that term, 
and define a level $2n$ approximation to the action to
be the one obtained by including fields up to level $n$ and terms in the
action up to level $2n$. 
Thus for example, a level 4 approximation to the action will involve
string field components up to level 2. This is the approximation we
shall be using to compute the action. 

The string field action has a gauge invariance which can be used
to choose a gauge in which 
\be
\label{egauge} 
b_0\hp=0, \qquad \xi_0\hp=0\, ,
\ee
which is valid at least at the linearized level. All relevant states
in the ``small'' Hilbert space
can be obtained by acting with ghost number zero combinations of
oscillators $\{b,c,\beta,\gamma, L^m, G^m\}$ on 
$|\widetilde\Omega\rangle$. The $b_0\hp=0$ gauge condition allows us to 
ignore states with a $c_0$ oscillator in them. The states one finds up to
$L_0$ eigenvalue $\frac{3}{2}$ are given in Table 1. For ease of notation we have not
included the CP factor. 
The string field $\Phi$ which uses the ``large'' Hilbert
space, is obtained by acting the states of the table with $\xi_0=\oint dz \frac{\xi(z)}{z}$.

As shown in \cite{bsz}, the string field theory action, restricted to ${\cal H}_1$ 
has a $Z_2$ twist symmetry under which string field components associated with a vertex operator of dimension $h$ carry charge $(-1)^{h+1}$ for even $2h$, and $(-1)^{h+{1\over2}}$ for 
odd $2h$. Thus it is possible to consistently restrict the string field $\wh{\Phi}$ to
be twist even.

\subsubsection*{States and Vertex Operators}
 The states that appear in the string field up to level two are 
\[
|\tilde{\Omega}\rangle \,,\,\{ c_{-1}\beta_{-\frac{1}{2}}\,,\,b_{-1}\gamma_{-\frac{1}{2}}\,,\,G_{-\frac{3}{2}}\}|\tilde{\Omega}\rangle \,,\,\{b_{-1}c_{-1}\,,\,\beta_{-\frac{1}{2}}\gamma_{-\frac{3}{2}}\,,\,\beta_{-\frac{3}{2}}\gamma_{-\frac{1}{2}}\,,\,(\beta_{-\frac{1}{2}}\gamma_{-\frac{1}{2}})^{2}\,,\,L_{-2}\}|\tilde{\Omega}\rangle\,,
\]
where $|\tilde{\Omega}\rangle $ is the tachyon state. The contribution to the 
tachyon potential of the tachyon and the 
three level $\fracs{{3}}{{2}}$ states was 
calculated in \cite{bsz}. We are interested in 
the contribution of level two states to the 
tachyon potential. The vertex operators
corresponding to these level two states are linear combination of the following five vertex operators
\[ 
\de^2c\, \ep \,, T_m \,c\, \ep \,, T_{\xi \eta} \,c\, \ep \,, T_{\phi}\,c\, \ep \,, \de^2 \phi\, c\, \ep \,.
\]
We can pass to the string field $\wh{\Phi}$ by acting the above vertex operators with
$\xi_0$. Denoting the tachyon vertex operator by $\wh{T}$, the three vertex operators at level ${3 \over 2}$
by $\wh{A}$, $\wh{E}$ and $\wh{F}$ and the five vertex operators at level 2 by $\wh{V}_\alpha$ 
($\alpha=1\dots 5$)
\begin{eqnarray} \label{e18} 
\nn \widehat T &=& \xi\, c \,e^{-\phi}\otimes \sigma_1 \,\,,\,\,\widehat A  =\xi\partial \xi \,c \,\partial^2 c\,e^{-2\phi} \otimes I \,\,,\,\,\widehat E = \xi\,\eta \otimes I\,\,,\,\,\widehat F =  \xi\, G_m\, c \,e^{-\phi} \otimes I  \\
\nn \wh{V}_1&=& \xi\, T_m\, c\, \ep \otimes \sigma_1\,,\,
\wh{V}_2= \xi \,\de^2 c\, \ep \otimes \sigma_1 \,,\,
\wh{V}_3= \xi\, T_{\xi \eta}\, c \,\ep \otimes \sigma_1\,,\, \\ \nn
\wh{V}_4&=& \xi \,T_\phi \,c\, \ep \otimes \sigma_1 \,,\,
\wh{V}_5= \xi \,\de^2 \phi\, c \,\ep \otimes \sigma_1 \,.
\end{eqnarray}
\begin{table}
\begin{eqnarray*} 
\begin{array}{|c|c|c|c|} \hline
\rule{0mm}{5mm}L_{0}          & \mbox{Level}       & GSO(+)       &GSO(-)                          \\ \hline
\rule{0mm}{5mm} -\frac{1}{2} &                 0        &                  &   |\tilde{\Omega}\rangle             \\  \hline
\rule{0mm}{5mm}~~0                & \frac{1}{2}                     & c_{0}\beta_{-\frac{1}{2}}|\tilde{\Omega}\rangle &     \\ \hline
\rule{0mm}{5mm} ~~ \frac{1}{2} & 1 & &\beta_{-\frac{1}{2}}\gamma_{-\frac{1}{2}}|\tilde{\Omega}\rangle   \\ \hline
\rule{0mm}{5mm}~~1 & \frac{3}{2} & \{c_{-1}\beta_{-\frac{1}{2}}\,,\,b_{-1}\gamma_{-\frac{1}{2}}\,,\,G_{-\frac{3}{2}}\}|\tilde{\Omega}\rangle &  \\ \hline
\rule{0mm}{5mm}~~\frac{3}{2} & 2 & &\{b_{-1}c_{-1}\,,\,\beta_{-\frac{1}{2}}\gamma_{-\frac{3}{2}}\,,
\,\beta_{-\frac{3}{2}}\gamma_{-\frac{1}{2}}\,, \\ & & & (\beta_{-\frac{1}{2}})^2 (\gamma_{-\frac{1}{2}})^2\,,L_{-2}\}|\tilde{\Omega}\rangle  \\ \hline
\end{array}
\end{eqnarray*} 
\end{table}
Therefore, the
general twist even string field up to level $\fracs{{3}}{{2}}$, satisfying the gauge
condition  \refb{egauge}, has the following form:
\be \label{ecc1}
\hp = t \,\wh T + a \,\wh A + e \,\wh E + f\, \wh F +v_1 \wh V_1\,+v_2 \wh V_2\,+v_3 \wh V_3\,
+ +v_4 \wh V_4\,+v_5 \wh V_5\,.
\ee
\subsubsection*{Expansion of the String Field Action}
We shall now substitute (\ref{ecc1}) into the action (\ref{e00}) and keep
terms to all orders in $t$, but only up to quadratic order in $a$, $e$,
$f$ and $v_\alpha$. Although the string field action contains vertices up to 
arbitrarily high order, at a given level, the action has a finite number
of terms (as shown in \cite{bsz}). The string field action, for $\wh{\Phi}$ restricted to 
${\cal H}_1$ and twist even fields is given by \cite{bsz}
\bea \label{e17}
S \hskip-5pt&&\hskip-10pt = {1\over 2g^2} \lllb 
\,\,{1\over 2} 
 \,(\hq \hp)\,\,(\he \hp)\, 
 + {1\over 3}
\,(\hq\hp)\,  \hp \,(\he\hp)  +
{1\over 12}  \,(\hq\hp) \, \Bigl( \hp^2\, 
(\he\hp) - \hp\,(\he\hp)\,\hp \, \Bigr) 
 \cr  
&& \qquad +{1\over 60}
 \,(\hq\hp) \, \Bigl( \hp^3\, (\he\hp) - 3\hp^2\,(\he\hp)\,\hp \Bigr) 
\cr
&& \qquad  + {1\over 360}\,(\hq\hp) \, 
\Bigl( \hp^4\, (\he\hp) - 4\hp^3\,(\he\hp)\,\hp
+ 3\hp^2\,(\he\hp)\,\hp^2  \Bigr) 
\rrrb \,. 
\eea
\subsection*{Calculation of the Tachyon Potential}
The calculation of the tachyon potential is simple but tedious. 
Using the expansion of $\hp$ to level 4 fields  
(\ref{ecc1}) in the string field action (\ref{e17}), we collect various
terms in the fields $t,\,a,\,e,\,f$ and $v_\alpha$. The coefficient of each
term is a sum of ``big'' correlators $\lll \dots \rrr$. We can then use the definition
of $\lll \dots \rrr$\footnote{We will need to conformal and BRST transformation properties 
of the various
operators appearing inside $\lll \cdots \rrr$. These are given in the appendix.}  in
(\ref{e2ff}), and trace over the $\sigma$ matrices to express each ``big'' correlator
in terms of ordinary correlators in the matter ghost CFT. These can then
be calculated 
using the normalization in (\ref{eb1}). We will give a simple example at
each level of the action to illustrate the procedure and give some useful
tricks which simplify the calculation. 
\normalsize
\subsubsection*{Level 2 terms}
The only non zero level 2 terms in the action involving $v_\alpha$'s
are of the form $t^3 v_\alpha$
\footnote{$tv_\alpha$ terms are also level 2 but the coefficients of these terms vanish.}
.
The relevant term in the string field action which gives rise to this term
is
\[
S_{quartic}=
 \fracs{{1}}{{2g^2}}\lll\fracs{{1}}{{12}}  \,(\hq\hp) \, \Bigl( \hp^2\, 
(\he\hp) - \hp\,(\he\hp)\,\hp \, \Bigr) \rrr \,.
\]
We can write $\hp=\Phi_{+} \otimes I + \Phi_{-}\otimes \sigma_1 $, $\hq= Q_B \otimes
\sigma_3$ and $\eta_0=\eta_0 \otimes \sigma_3$. Tracing over the $\sigma$ matrices,
we obtain
\bea
-12g^2 S_{quartic} &=& \lll (Q_B \Phi_{-}) \Phi_{-}^2(\eta_0 \Phi_{-}) \rrr+\lll
(Q_B \Phi_{-}) \Phi_{-}(\eta_0 \Phi_{-})\Phi_{-} \rrr+\dots
\label{quar}
\eea
where $\dots$ represents other terms in $S_{quartic}$ which are irrelevant for
the calculation of the coefficient of $t^3 v_\alpha$ (they are also irrelevant for 
$t^2 v_\alpha v_\beta$ which are level 4 terms). Here
\be
\Phi_{-}=t{T}+v_1 {V}_1 + v_2 {V}_2 + v_3 {V}_3 +v_4 {V}_4+v_5 {V}_5
\label{exp}\,.
\ee
Substituting (\ref{exp}) in (\ref{quar}) and collecting the $t^3 v_\alpha$ terms together, 
we find \footnotesize{
\bea
-12 g^2 S  =  \cdots & +&t^3 v_\alpha \Bigl(\, \lll \,( Q_B T) \, T\, T (\eta_0 V_\alpha) \, \rrr\,+ \, \lll \, (Q_B T) \, T\, V_\alpha (\eta_0 T) \, \rrr\,+\, \lll \,( Q_B T) \, V_\alpha\, T (\eta_0 T) \, \rrr\,\nn \\ &+&\, \lll \, (Q_B V_\alpha) \, T\, T (\eta_0 T) \, \rrr\,+\, \lll \, (Q_B T) \, T\,  (\eta_0 V_\alpha) T\, \rrr\,+ \, \lll \, (Q_B T) \, T\,  (\eta_0 T) V_\alpha\, \rrr\,\nn \\ &  +&\, \lll \, (Q_B T) \, V_\alpha\,  (\eta_0 T) T\, \rrr\,+\, \lll \, (Q_B V_\alpha) \, T\,  (\eta_0 T) T\, \rrr\, \Bigr) \cdots
\label{quarexp}
\eea} \normalsize
To calculate the $t^3 v_\alpha$ term, we only need to calculate one correlator at
general points: 
$\,  \lll (Q_2 T) \, T\, V_\alpha \, (\eta_0 T) \, \rrr$. All other correlators 
which appear in (\ref{quarexp}) can be related to this one by using $\lll Q_B(\dots) \rrr=0$
and $\lll \eta_0(\dots)\rrr=0$ as we now show. 
We define
 \footnote{We will drop the $n$ index from $f_k^{(n)}$ for clarity. In what follows
in the calculation of level 2 terms, it is assumed that $n=4$.}
\footnotesize{
\bea
X(I,J,K,L)& \equiv & \lll\bigl(Q_2 T(w_I)\bigr) \, T (w_J) \, V_\alpha(w_K) \, \bigl(\eta_0 T(w_L)\bigr) \rrr \nn \\
& \equiv & \frac{(f'_{K})^{\frac{3}{2}}}{(f'_{I}f'_{J}f'_{L})^{\frac{1}{2}}}\langle\bigl( Q_2 T(w_I)\bigr)\, T(w_J)
\, \Bigl(V_\alpha(w_K)  +{\cal P}_{V_\alpha}^K U_\alpha(w_K) +{\cal R}_{V_\alpha}^KT(w_K)\Bigr) \bigl(\eta_0 T(w_L)\bigr) \rangle \label{X},
\eea
}
\normalsize
where ${\cal P}_{V_\alpha}^K$ and ${\cal R}_{V_\alpha}^K$ are defined in the appendix, 
$U_\alpha$ is the operator multiplying  ${\cal P}_{V_\alpha}^K$ in the conformal 
transformation of $V_\alpha$ ($U_1=0,\, U_2= \xi \de c \ep, \, U_3= \de \xi c \ep, \, U_4=U_5=
\xi \de \phi c \ep$). 
Notice that $X(1,2,3,4)=\lll (Q_B T) \, T\, V_\alpha (\eta_0 T) \rrr$ where we have defined
$w_1=1, \, w_2=i, \, w_3=-1, \, w_4=-i,$ and $i=\sqrt{-1}$.
\normalsize
Using $\lll \eta_0((Q_BT)\,T\,V_\alpha \, T) \rrr=0$ we obtain\footnote{We have used $\eta_0 Q_2 T=
-Q_2 \eta_0 T=0$.}
\footnotesize{
\begin{eqnarray*}
\lll\bigl( Q_2 T(w_{I})\bigr) \, T(w_{J}) \, \bigl(\eta_0V(w_{K})\bigr) \, T(w_{L}) \rrr&=& \lll \bigl(Q_2 T(w_{I})\bigr) \bigl(\eta_0 T(w_{J})\bigr) 
V_\alpha(w_{K}) T(w_{L}) \rrr\\&& \hspace{1.1in} + \lll \bigl(Q_2 T(w_{I})\bigr) T(w_{J}) V_\alpha(w_{K}) \bigl(\eta_0 T(w_{L})\bigr) \rrr,\\ & = & -X(I,L,K,J)+X(I,J,K,L). 
\end{eqnarray*}} \normalsize
Similarly, using $ \lll \, Q_2 \bigl(V_\alpha \, T T (\eta_0 T)\bigr) \, \rrr=0$ we obtain
\footnotesize{
\[
\lll \bigl(Q_2 V_\alpha(w_I)\bigr) T(w_J) T(w_K) \bigl(\eta_0 T(w_L)\bigr) \rrr=X(K,J,I,L)-X(J,K,I,L).
\]}\normalsize
Hence we can write (\ref{quarexp}) in terms of $X(I,J,K,L)$ \footnote{$S_{t^3v_\alpha}$ is the
coefficient of the $t^3 v_\alpha$ term in the action.}:
\footnotesize{
\bea
-12g^2 S_{t^3v_\alpha}&=& X(1,2,3,4)- X(1,3,4,2)+ X(1,2,4,3)+X(1,2,3,4)- X(1,3,2,4)-X(1,4,3,2)
\nn \\
& & -X(1,4,2,3)+X(1,2,4,3)+X(3,2,1,4)-X(2,3,1,4)+X(4,2,1,3) \nn \\
&& \, \,-X(2,4,1,3). \label{expx}
\eea}\normalsize
We will give some explicit results to calculate $X(I,J,K,L)$ for the case $t^3 v_2$. The
various CFT correlators appearing in (\ref{X}) for $\alpha=2$ are:
\footnotesize{
\begin{eqnarray*}
\langle \bigl(Q_{2}T(w_{I})\bigr)T(w_{J})V_{2}(w_{K})\bigl(\eta_{0}T(w_{L})\bigr)\,\rangle &=&-2\frac{w_{IL}}{w_{KL}},\\ \nn
\langle \bigl(Q_{2}T(w_{I})\bigr)T(w_{J})\xi \partial c e^{-\phi}(w_{K})\bigl(\eta_{0}T(w_{L})\bigr)\,\rangle &=&\frac{w_{IL}}{w_{KL}}(w_{JK}-w_{KL}),\\ \nn
\langle \bigl(Q_{2}T(w_{I})\bigr)T(w_{J})T(w_{K})\bigl(\eta_{0}T(w_{L})\bigr)\,\rangle &=&w_{JK}w_{IL}.\\ \nn
\end{eqnarray*}} \normalsize
Here $w_{IJ}=w_{I}-w_{J}$ etc. Using these expressions in (\ref{X}) and
substituting in (\ref{expx}), we find
\[ 
g^2 S=\cdots \,\,\,-\fracs{{17}}{{6}}\,t^3 v_2 \,\,\,\cdots
\]

\normalsize
\subsubsection*{Level 4}
At level 4, the non-zero terms are of the form $v_\alpha v_\beta$ and $t^2 v_\alpha v_\beta$.
As an example, we will discuss the $v_4 v_5$ term. 
\bea
2g^2 S_{v_\alpha v_\beta}&=&\fracs{{1}}{{2}} \Bigl(\, \lll (\hq \wh{V}_\alpha)\, (\he \wh{V}_\beta) \rrr + \lll (\hq \wh{V}_\beta) \, (\he \wh{V}_\alpha) \rrr \,\Bigr), \nn \\
& = & \fracs{{1}}{{2}} {\Tr}(\sigma_3 \sigma_1 \sigma_3 \sigma_1)\Bigl( \, \lll (Q_B V_\alpha)\, (\eta_0 V_\beta)  \rrr + \lll (Q_B V_\beta) \,(\eta_0 V_\alpha)  \rrr \,\Bigr), \nn \\
& = & -\Bigl(\, \lll (Q_B V_\alpha)\, (\eta_0 V_\beta)  \rrr+ \lll (Q_B V_\beta) \, (\eta_0 V_\alpha) \rrr\, \Bigr).
\eea
$\lll\, (Q_B V_\beta) \, (\eta_0 V_\alpha)\, \rrr$ can be related to $\lll \,(Q_B V_\alpha) \, (\eta_0 V_\beta)\, \rrr$ as follows\footnote{We have used
$Q_B\bigl(V_\alpha \,  (\eta_0 V_\beta)\bigr)=(Q_B V_\alpha) \, (\eta_0 V_\beta)-V_\alpha \, (Q_B \eta_0 V_\beta)$ which is different from (\ref{epr2a}). $V_\alpha$, being in the GSO($-$) sector is Grassman odd and
we had multiplied it by $\sigma_1$ in section which resulted in (\ref{epr2a}). Here, we have traced over the 
$ \sigma $  matrices.}
\[
0=\lll Q_B \bigl( V_\beta \,  (\eta_0 V_\alpha)\bigr) \rrr = \lll \,(Q_B V_\beta) \, (\eta_0 V_\alpha)\, \rrr -\lll V_\beta\, (Q_B \eta_0 V_\alpha)\, \rrr= \lll \,(Q_B V_\beta) \, (\eta_0 V_\alpha)\, \rrr +\lll V_\beta \, (\eta_0 Q_B V_\alpha)\, \rrr,
\]
\[
0=\lll\, \eta_0 \bigl( V_\beta\, (Q_B V_\alpha)\bigr) \,\rrr =  \lll (\eta_0 V_\beta) \, (Q_B V_\alpha) \rrr - \lll V_\beta \, (\eta_0 Q_B V_\alpha)\, \rrr,
\]
which implies
\bea
-2g^2 S_{v_\alpha v_\beta}&=& \lll \,(Q_B V_\alpha)\, (\eta_0 V_\beta)\,  \rrr -  \lll \,(\eta_0 V_\beta)\,( Q_B V_\alpha)\,  \rrr, \nn \\ & = & Y(1,2)-Y(2,1) \nn.
\eea
\footnotesize{
\bea
Y(I,J) &\equiv& \lll\,\bigl( Q_B V_\alpha(w_I)\bigr)\, \bigl(\eta_0 V_\beta(w_J)\bigr) \rrr, \nn \\  & \equiv & (f_i^{'})^{3/2}(f_j^{'})^{3/2} \langle\, Q_0\bigl(V_\alpha(w_I) + {\cal P}_{V_\alpha}^I U_\alpha(w_I) +{\cal R}_{V_{\alpha}^{I}}T(w_I)\bigr) \nn \\ & & \hspace{1.5in}
 \eta_0 \bigl(V_\beta(w_J) + {\cal P}_{V_\beta}^J U_\beta(w_J) +{\cal R}_{V_{\alpha}^{J}}T(w_J)\bigr)\rangle. \eea
}\normalsize
${\cal P}_{V_{\alpha}}^I$ and ${\cal R}_{V_{\alpha}}^I$ are defined in the appendix
\footnote{We have suppressed indices corresponding to $n$, the number of vertex operators
inside $\lll \dots \rrr$. For this subsection, $n=2$.}.  
Here, we have used $\phi$  momentum conservation to replace $Q_B$ by $Q_0$ since this is the
only part of $Q_B$ which contributes in the correlators. 

\begin{tabular}{l l}
$\langle \,\bigl(Q_0 V_4(w_I)\bigr) \bigl(\eta_0V_5(w_J)\bigr)\,\rangle =  \frac{15}{4}\frac{1}{(w_I-w_I)^3},$ &
$\langle\,\bigl(Q_0 \xi c \de \phi e^{-\phi} (w_I)\bigr)\,\bigl( \eta_0 V_5(w_J)\bigr)\,\rangle  = \frac{1}{2}\frac{1}{(w_I-w_J)^2},$ \\ &\\
$\langle\,\bigl(Q_0 T(w_I)\bigr)\,\bigl( \eta_0 V_5(w_J)\bigr)\rangle  = -\frac{1}{2}\frac{1}{w_I-w_J},$  &
$\langle \, \bigl(Q_0 V_4(w_I)\bigr)\, \bigl(\eta_0\xi c \de \phi e^{-\phi} (w_J)\bigr)\,\rangle =  \frac{1}{4}\frac{1}{(w_I-w_J)^2},$ \\&\\
$\langle \,\bigl(Q_0 V_4(w_I)\bigr)\,\bigl( \eta_0T(w_J)\bigr)\,\rangle =  \frac{1}{4}\frac{1}{w_I-w_J}, $&
$\langle\,\bigl(Q_0 \xi c \de \phi e^{-\phi} (w_I)\bigr)\,\bigl( \eta_0\xi c \de \phi e^{-\phi} (w_J) \bigr)\,\rangle  = 0,$  \\&\\
$\langle\,\bigl(Q_0 \xi c \de \phi e^{-\phi} (w_I)\bigr) \,\bigl(\eta_0T (w_J)\bigr)\, \rangle  =  \frac{1}{2},$ &
$\langle\,\bigl(Q_0 T(w_I)\bigr)\,\bigl( \eta_0 \xi c \de \phi e^{-\phi} (w_J)\bigr)\, \rangle  = -\frac{1}{2}, $\\ &\\ 
$\langle\, \bigl(Q_0 T(w_I)\bigr)\,\bigl( \eta_0 T(w_J)\bigr)\, \rangle = -\frac{1}{2}(w_I-w_J).$&\\ &\\
\end{tabular}
\newline
\newline
\normalsize
Setting $w_1=1$ and $w_2=-1$, we find
\[
-2g^2S_{v_4 v_5} = Y(1,2)-Y(2,1)=3+3=6.
\]

\subsection*{Level $\fracs{{7}}{{2}}$}
At this level we have terms of the form $t a v_{\alpha}$,
$tev_{\alpha}$, $tfv_{\alpha}$ and $t^{3}\,e\,v_{\alpha}$. The
coefficients of the terms of the form $t^{3}\,a\,v_{\alpha}$ and
$t^{3}\,f\,v_{\alpha}$ vanish because of $\phi$ momentum conservation.
Here we will explain the calculation of the coefficient of the cubic
term $tav_{1}$.  The relevant term in the string field action which
gives rise to cubic terms is
\[
S_{cubic}=\fracs{{1}}{{2g^{2}}}\lll\, \fracs{{1}}{{3}}(\widehat{Q}_{B}\widehat{\Phi})\widehat{\Phi}(\widehat{\eta}_{0}\widehat{\Phi})\,\rrr.
\]
In terms of GSO(+) ($\Phi_{+}$) and GSO($-$) ($\Phi_{-}$)
terms in the string field action, the cubic part of the action after
 tracing over the $\sigma$ matrices becomes
\be
S_{cubic}=-\fracs{{1}}{{3g^{2}}}\Bigl \{(Q_{B}\Phi_{+})\Phi_{-}(\eta_{0}\Phi_{-})+(Q_{B}\Phi_{-})\Phi_{+}(\eta_{0}\Phi_{-}) \Bigr \}+\cdots 
\label{cubic}
\ee
where $(\cdots)$ represent terms which are not relevant for the calculation of the coefficient of $tav_{1}$. The expansion of $\Phi_{+}$ and $\Phi_{-}$ relevant here is
\[
\Phi_{+}=a\,A\,,\,\,\,\Phi_{-}=t\,T+v_{1}\,V_{1}\,.
\] 
Substituting the above expansion in (\ref{cubic}) we obtain,  
\begin{eqnarray}
S|_{tav_{1}}=-\fracs{{1}}{{3g^{2}}}\{\lll\, (Q_{B}T) \,A\,(\eta_{0}V_{1})\,\rrr \nn  
+\lll\, (Q_{B}A) \,T\,(\eta_{0}V_{1})\,\rrr \nn
 -\lll\, (Q_{B}V_{1}) \,T\,(\eta_{0}A)\,\rrr \\ \nn
 -\lll\, (Q_{B}T) \,V_{1}\,(\eta_{0}A)\,\rrr  \nn
 +\lll\, (Q_{B}A) \,V_{1}\,(\eta_{0} T)\,\rrr \nn
 +\lll\, (Q_{B}V_{1}) \,A\,(\eta_{0}T)\,\rrr\} \nn
\end{eqnarray}
To calculate the coefficient of $tav_{1}$ term we actually only need to calculate one correlator. 
This is partly due to the fact that in the correlators given above involving $V_{1}$  we can
replace $V_{1}$ by a multiple of $T$. \footnotesize
\begin{eqnarray*}\nn
\lll\, \bigl(Q_{B}V_{1}(w_{I})\bigr)A(w_{J})\bigl(\eta_{0}T(w_{K})\bigr)\,\rrr &=&(f_{I}')^{\fracs{{3}}{{2}}}(f_{J}')(f_{K}')^{-\fracs{{1}}{{2}}} \langle \,Q_{B}(V_{1}(w_{I})+{\cal R}^{I}_{V_{1}}T(w_{I})) \nn \\ & &
\hspace{1.5in}  (A(w_{J})+{\cal P}^{J}_{A}\,U_{A}(w_{J}))\eta_{0}(T(w_{K}))\,\rangle, \\ \nn
&=&(f_{I}')^{2}\,{\cal R}^{I}\lll\,\bigl( Q_{B}T(w_{I})\bigr) \,A(w_{J})\,\bigl(\eta_{0}T(w_{K})\bigr)\,\rrr, \\&\equiv&  (f_{I}')^{2}\,{\cal R}^{I}\,C(I,J,K),\\ \nn
\end{eqnarray*}\normalsize
where in the second line we have used the fact that since neither $A$ nor $T$ have fields from the matter sector therefore
the term with $V_{1}$ in the correlator
vanishes. Similarly  \footnote{In this subsection we will suppress some of
the indices on ${\cal R}^{I,3}_{V_{1}}\equiv {\cal R}^{I}$.}
\begin{eqnarray*}\nn
\lll \, \bigl(Q_{B}T(w_{I})\bigr) \,A(w_{J})\,\bigl(\eta_{0}V_{1}(w_{K})\bigr)\,\rrr &=& (f_{K}')^{2}{\cal R}^{K} \,C(I,J,K), \nn
\end{eqnarray*}\normalsize
where the factor ${\cal R}^{K}$ comes from the conformal
transformation of $V_{1}$ and $(f'_{K})^{2}$ is needed to compensate
for the difference in the conformal dimensions of $T$ and $V_{1}$. 
Using the fact that $Q_{2}\eta_{0}(T)=Q_{2}\eta_{0}(V_{1})=0$ and $Q_{B}$ acts like
a derivation it follows that, 
\begin{eqnarray*}
\lll\, \bigl(Q_{B}A(w_{I})\bigr)\,V_{1}(w_{J})\,\bigl(\eta_{0}T(w_{K})\bigr)\,\rrr &=&-\lll \,\bigl(Q_{B}V_{1}(w_{J}\bigr)\,A(w_{I})\,\bigl(\eta_{0}T(w_{K})\bigr)\,\rrr, \\ \nn
 \lll\, \bigl(Q_{B}T(w_{I})\bigr) \,V_{1}(w_{J})\,\bigl(\eta_{0}A(w_{K})\bigr)\,\rrr&=&\lll \,\bigl(Q_{B}T(w_{I})\bigr) \,A(w_{K})\,\bigl(\eta_{0}V_{1}(w_{J})\bigr)\,\rrr,    \\ \nn
\lll\, \bigl(Q_{B}A(w_{I})\bigr) \,T(w_{J})\,\bigl(\eta_{0}V_{1}(w_{K})\bigr)\,\rrr&=&-\lll \,\bigl(Q_{B}T(w_{J})\bigr) \,A(w_{I})\,\bigl(\eta_{0}V_{1}(w_{K})\bigr)\,\rrr.
\end{eqnarray*}
Thus using the properties of $Q_{B}$ and $\eta_{0}$ we have reduced
the calculation to one correlator 
\begin{eqnarray}\nn
g^{2}S|_{tav_{1}}=&-&\fracs{{1}}{{3}}\{ (f_{3}')^{2}\,{\cal R}^{3}\,C(1,2,3)-(f_{3}')^{2}\,{\cal R}^{3}\,C(2,1,3)-(f'_{1})^{2}{\cal R}^{1}C(1,3,2) \\ \nn
&-&(f'_{2})^{2}{\cal R}^{2}C(1,3,2)
-(f'_{2})^{2}{\cal R}^{2}C(2,1,3)+(f'_{1})^{2}{\cal R}^{1}C(1,2,3)\}.\nn
\end{eqnarray}
\begin{eqnarray}\nn
C(I,J,K)&=&\lll \,\bigl(Q_{B}T(w_{I})\bigr) \,A(w_{J})\,\bigl(\eta_{0}T(w_{K})\bigr)\,\rrr\\, \nn
&=&(f'_{I})^{-\frac{1}{2}}(f'_{J})(f'_{K})^{-\frac{1}{2}}\langle \,\bigl(Q_{B}T(w_{I})\bigr)[A(w_{J})+{\cal P}_{A}^{J}\xi\partial\xi c\partial c e^{-2\phi}(w_{J})]\,\bigl(\eta_{0}T(w_{K})\bigr)\rangle, \\ \nn
&=&(f'_{I})^{-\frac{1}{2}}(f'_{J})(f'_{K})^{-\frac{1}{2}}(-2\frac{w_{IK}}{w_{JK}}-{\cal P}^{J}_{A}w_{IK}).
\end{eqnarray}\normalsize
\[
g^{2}S|_{tav_{1}}=-\fracs{{50}}{{18}}.
\]
\subsection*{Tachyon Potential} 
We now give the result of our calculation \footnote{Exact coefficients of the terms $t^3ev_{\alpha}$ are sum of 40 correlators and have very long expressions in terms of radicals, therefore only their decimal expansion is given.}.
\begin{figure}
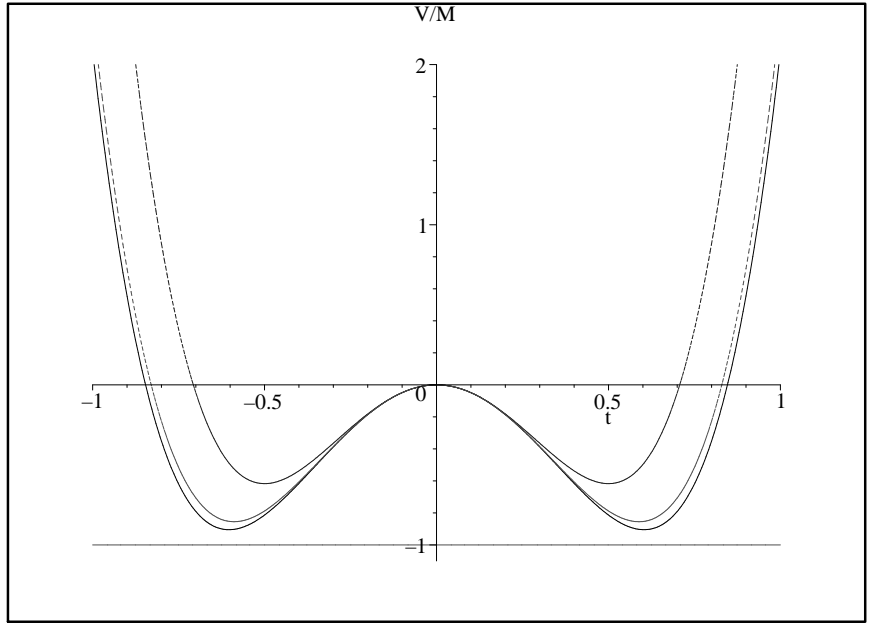

\PSbox{figure0101.eps hoffset=30 voffset=280 hscale=50 vscale=50 angle=-90}{4.8in}{3.4in}
\caption{The effective tachyon potential in the three approximations.} 
\end{figure}
\begin{eqnarray*}
g^2 S_{0}&=& \fracs{{1}}{{4}}\,t^2 -
\fracs{{1}}{{2}}\,t^4, \\ \nn
g^2 S_{\fracs{{3}}{{2}}}&=& -a\,t^2 -  
\fracs{{1}}{{4}}\, e\,t^2 
- \fracs{{5}}{{96}} \sqrt{50 + 22 \sqrt{5}}
e\,t^4, \\ \nn
g^2S_{2}&=&{\displaystyle \fracs{{15}}{{4}}} \,{t^3v_1}
 - {\displaystyle \fracs{{17}}{{6}}} \,{t^3v_2} - {\displaystyle 
\fracs{{1}}{{4}}} \,{t^3v_3} + {\displaystyle \fracs{{35}}{{12}}} \,
{t^3v_4} - {\displaystyle \fracs{{11}}{{6}}} \,{t^3v_5}, \\ \nn
g^2 S_{3}&=&2\,a\,e + 5\,f^2-\Bigl(\,\fracs{{1}}{{\sqrt 2}} - \fracs{{1}}{{24}}\,\Bigr)\, e^2\,t^2  
+ \fracs{{5}}{{18}}  \,e^2 \,t^4
+\, \fracs{{5}}{{4}} ( 4\sqrt{2}\,-1) \,f^2\,t^2  \\ \nn
&+& \fracs{{1}}{{12}}
\,(\, 3 + 40 \sqrt{2}\,) \,ae\,t^2 - \fracs{{5}}{{12}}\, (\,10 \sqrt 2 - 1\,)
ef\,t^2\,, \\ \nn
g^2 S_{\fracs{{7}}{{2}}}&=&\mbox{}  - {\displaystyle \fracs{{25}}{{9}}} \,{t\,a\,v_1} + {\displaystyle \fracs{{44}}{{27}}} \,{t\,a\,v_2} + 
{\displaystyle \fracs{{44}}{{27}}} \,{t\,a\,v_3} - {\displaystyle 
\fracs{{329}}{{27}}} \,{t\,a\,v_4} + {\displaystyle \fracs{{64}}{{9}}} \,
{t\,a\,v_5} \\
 &- & \mbox{}   {\displaystyle \fracs{{25}}{{36}}} \,
{t\,e\,v_1} + {\displaystyle \fracs{{35}}{{27}}} \,{t\,e\,v_2} - 
{\displaystyle \fracs{{17}}{{18}}} \,{t\,e\,v_3} - {\displaystyle 
\fracs{{41}}{{108}}} \,{t\,e\,v_4} + {\displaystyle \fracs{{8}}{{27}}} \,
{t\,e\,v_5} \\
 &- & \mbox{}   {\displaystyle \fracs{{80}}{{9}}} \,{t\,f\,v_1} 
- {\displaystyle \fracs{{80}}{{27}}} \,{t\,f\,v_3} + 
{\displaystyle \fracs{{160}}{{9}}} \,{t\,f\,v_4} - {\displaystyle 
\fracs{{320}}{{27}}} \,{t\,f\,v_5} \nn \\ 
&+&4.35732\,{t^3 \, e \, v_1} - 4.77364\,
{t^3 \, e \, v_2} + .605194\,{t^3 \, e \, v_3} + 4.351715\,{t^3 \, e \, v_4} \nn \\ &-&  
1.94348\,{t^3 \, e \, v_5}, \nn \\
g^2 S_{4}&=& -  \fracs{{45}}{{8}} \,{v_1}^{2} + 3\,
{v_2}^{2} +  \fracs{{3}}{{4}} \,{v_3}^{2}
 -  \fracs{{15}}{{8}} \,{v_4}^{2} - 
 \fracs{{3}}{{2}} \,{v_5}^{2} + 3\,{v_4}\,
{v_5}  \nn\\&-&   {\displaystyle \fracs{{945}}{{64}}} \,t^2{v_1}^{2}
 - {\displaystyle \fracs{{35}}{{6}}} \,
t^2 {v_2}^{2} + {\displaystyle \fracs{{5}}{{4}}} \,t^2{v_3}^{2}
 \mbox{} - {\displaystyle \fracs{{3511}}{{192}}} t^2\,{v_4}^{2
} - {\displaystyle \fracs{{161}}{{24}}} \,t^2{v_5}^{2}\\&+&  
{\displaystyle \fracs{{255}}{{16}}} \,t^2{v_1}\,{v_2} + 
{\displaystyle \fracs{{45}}{{32}}} \,t^2{v_1}\,{v_3} - 
{\displaystyle \fracs{{525}}{{32}}} \,t^2{v_1}\,{v_4} + 
{\displaystyle \fracs{{165}}{{16}}} \,t^2{v_1}\,{v_5} + 
{\displaystyle \fracs{{19}}{{48}}} \,t^2{v_2}\,{v_3} \\
 &+ & \mbox{}  {\displaystyle \fracs{{991}}{{48}}} t^2\,{v_2}\,
{v_4} - {\displaystyle \fracs{{143}}{{12}}} \,t^2{v_2}\,
{v_5} - {\displaystyle \fracs{{65}}{{32}}} \,t^2{v_3}\,
{v_4} + {\displaystyle \fracs{{7}}{{8}}} \,t^2{v_3}\,
{v_5} + {\displaystyle \fracs{{985}}{{48}}} \,t^2{v_4}\,
{v_5}. \nn \\
\end{eqnarray*}
The potential is given by
\[
V(t,a,e,f,v_{i})=-S(t,a,e,f,v_i)=-2\pi^{2}M\, g^2 S(t,a,e,f,v_{i})\,,
\]
where $M=\frac{1}{2\pi^2 g^2}$. The potential has a minimum
for the following values of $t,a,e,f,v_i$. 
\begin{eqnarray*}
&&t=\pm 0.603455, \,\,\,a=0.059291, \, \, \,e=0.030980, f=0.015746,\,\,\,v_1=\pm0.035688, \,\,\,\\&&
v_2=\pm0.0571190, \,\,\,  v_3=\pm0.0153737,\,\,\, v_4=\pm0.031538,\,\,\, v_5=\mp 0.006012.
\end{eqnarray*}
The value of the potential at the minimum is
\[
V=-0.90454 \,M\,.
\]
The expected answer for the value of the potential at the minimum
is $-M$. Thus we see that the level four approximation produces
 $90.5\%$ of the expected value.  
We can integrate out the fields $a, \, e, \, f,\,$ and $ v_\alpha$
to obtain an effective potential for the tachyon which
is given by
\[
V_{effective}=-2 \pi^{2}M\frac{P(t)}{Q(t)},
\]
where
\begin{eqnarray*}\nn
P(t)&=& 0.8168083768\,10^{-5}\,t^{2}+0.0002047077013\,t^{4} 
+0.002821258794\,t^{6}\\ \nn
&+& 0.02594499700\,t^{8} 
+0.1738511409\,t^{10} 
+ 0.8795191385\,t^{12}\\ \nn
 &+& 3.409051376\,t^{14}
+10.11847495\,t^{16}
+ 22.58030878\,t^{18}\\ \nn
&+& 35.88635644\,t^{20}
+ 33.50595320\,t^{22}
 - 4.683061411\,t^{24} \\ \nn
&-& 77.91238815\,t^{26}
-149.1412411\,t^{28}
- 168.7145393\,t^{30} \\ \nn
 &-& 125.0599362\,t^{32} 
- 57.59279282\,t^{34} 
-12.56479811\,t^{36},
\end{eqnarray*}
and
\begin{eqnarray*}\nn
\lefteqn{Q(t)= 0.00003267233507 +
0.0008678393076\,t^{2} +0.01254363691\,t^{4}}\\ \nn
&+&0.1215742201\,t^{6}+0.8648433845\,t^{8}+4.706706334\,t^{10}\\ \nn
&+& 20.03806794\,t^{12}+67.54526560\,t^{14} + 181.3752555\,t^{16}\\ \nn
&+& 388.8655996\,t^{18}+ 664.8883542\,t^{20}+ 902.1103038\,t^{22}\\ \nn
&+& 959.2815429\,t^{24}+778.5205080\,t^{26} + 459.3896684\,t^{28}\\ \nn
&+& 177.3949239\,t^{30}+ 33.42209885\,t^{32}.
\end{eqnarray*}

%
\section*{Acknowledgements}
We would like to thank Barton Zwiebach and Leonardo Rastelli for valuable discussions
and P. De Smet and J. Raemaekers for useful correspondence.
This research was supported in part by the US Department of Energy
under contract \#DE-FC02-94ER40818.

\section{Appendix}
{\bf Conformal Transformations:}
\begin{eqnarray} \nn
f\cdot T(0)&=&(f'(0))^{-\frac{1}{2}}\,T(w)\\ \nn
f\cdot A(0)&=&f'(0)\,\{ A(w)-\fracs{{f''(0)}}{{(f'(0))^{2}}}\,c\partial c\,\xi\,\partial\xi e^{-2\phi}(w)\,\}\\ \nn
f\cdot E(0)&=&f'(0)\,\{E(w)-\fracs{{f''(0)}}{{2(f'(0))^{2}}} \,\}\\ \nn
f\cdot F(0)&=&f'(0)\,F(w)\,\\ \nn
f\cdot V_{1}(0)&=&(f'(0))^{\frac{3}{2}}\,\{ V_{1}(w)-\fracs{{15}}{{12}} \{z,f\}\xi\,c\,e^{-\phi}(w)\}\\ \nn
f \cdot V_{2}(0) &=& (f'(0))^{\frac{3}{2}}\,\{V_{2}(w)-\fracs{{f''(0)}}{{(f'(0))^{2}}}\xi \partial c\,e^{-\phi}(w)+(2\frac{{(f''(0))^{2}}}{{(f'(0))^{4}}}-\frac{{f'''(0)}}{{(f'(0))^{3}}})\xi c e^{-\phi}(w)\,\} \nn \\
f\cdot V_{3}(0)&=&  (f'(0))^{\frac{3}{2}}\,\{ V_{3}(w)+\fracs{{1}}{{6}}\,\{z,f\}\,\xi\,c\,e^{-\phi}(w)+
\frac{f''(0)}{2(f'(0))^2}\de \xi c \ep \, \} \\ \nn
f \cdot V_{4}(0)&=& (f'(0))^{\frac{3}{2}}\,\{ V_{4}(w)-\fracs{{1}}{{2}}\fracs{{f''(0)}}{{(f'(0))^{2}}}\xi c\partial \phi e^{-\phi}+(\fracs{{1}}{{12}}\fracs{{f'''(0)}}{{(f'(0))^{3}}}-\fracs{{2}}{{3}}\{z,f\})\,\xi\,c\,e^{-\phi}(w)\,\} \\ \nn
f \cdot V_{5}(0)&=&(f'(0))^{\frac{3}{2}}\,\{ V_{5}(w)+\fracs{{f''(0)}}{{(f'(0))^{2}}}\xi c\partial \phi e^{-\phi}
+(\fracs{{1}}{{2}}\{z,f\} -\fracs{{1}}{{6}}\fracs{{f'''(0)}}{{(f'(0))^{3}}})\xi c e^{-\phi} \}\nn
\end{eqnarray}
\begin{eqnarray}
{\cal P}^{I}_{A}&=&-\fracs{{f''_{I}(0)}}{{(f'_{I}(0))^{2}}}\,,\,{\cal P}_{E}^{I}=-\fracs{{f''_{I}(0)}}{{2(f'_{I}(0))^{2}}}\,,\, {\cal P}_{F}^{I}=0\\ \nn \\
{\cal P}^{I}_{V_{1}}&=&0 \,,\,{\cal P}^{I}_{V_{2}}= -\fracs{{f''_{I}(0)}}{{(f'_{I}(0))^{2}}}\,,\,{\cal P}^{I}_{V_{3}}=\fracs{{f''_{I}(0)}}{{2(f'_{I}(0))^{2}}} \,,\,{\cal P}^{I}_{V_{4}}= -\fracs{{f''_{I}(0)}}{{2(f'_{I}(0))^{2}}}\,,\,{\cal P}^{I}_{V_{5}}=\fracs{{f''_{I}(0)}}{{(f'_{I}(0))^{2}}} \,,\, \\ \nn \\
{\cal R}^{I}_{V_{1}}&=&-\fracs{{15}}{{12}}\{z,f\}\,,\,{\cal R}^{I}_{V_{2}}=2\frac{{(f''(0))^{2}}}{{(f'(0))^{4}}}-\frac{{f'''(0)}}{{(f'(0))^{3}}} \,,\,{\cal R}^{I}_{V_{3}}=\fracs{{1}}{{6}}\{z,f\}\,,\\ \nn \\
{\cal R}^{I}_{V_{4}}&=&\fracs{{1}}{{12}}\fracs{{f'''(0)}}{{(f'(0))^{3}}}-\fracs{{2}}{{3}}\{z,f\} \,,\,{\cal R}^{I}_{V_{5}}= -\fracs{{1}}{{6}}\fracs{{f'''(0)}}{{(f'(0))^{3}}}+\fracs{{1}}{{2}}\{z,f\}  \,,\,
\end{eqnarray}

{\bf BRST transformations:}
\bea 
\nn Q_B = \oint dz j_B(z)
= \oint dz \Bigl\{  c \bigl( T_m + T_{\xi\eta} + T_\phi)
+ c \partial c b +\eta \,e^\phi
\, G_m - \eta\partial \eta e^{2\phi} b \Bigr\}\, ,
\eea
\begin{eqnarray}\nn
Q_{0} &\equiv& \oint dz \Bigl\{  c \bigl( T_m + T_{\xi\eta} + T_\phi)
+ c \partial c b, \\ \nn
Q_{1} &\equiv& \oint dz \,\eta e^{\phi}G_{m}  \,\,\,,\,\,\,
Q_{2} \equiv  - \oint dz \,\eta\partial \eta e^{2\phi} b. \nn  
\end{eqnarray}
\[
U_2=\xi \de c \ep,\,\,\,U_3=\de \xi c \ep, \,\,\,U_4=U_5=\xi c \de \phi \ep.
\]
{\bf $Q_{0}$:}
\begin{eqnarray*}\nn
Q_{0}(T)&=&-\fracs{{1}}{{2}}\xi c\partial c e^{-\phi},\\
Q_{0}(A)&=&-2c\partial c \partial^{2}c \xi \partial \xi e^{-2\phi},\\ 
Q_{B}(E)&=&c \de \xi \eta- \de c \eta \xi-c \de \eta \xi-\fracs{{1}}{{2}} \de^2 c +\eta e^{\phi}G_{m}+2\partial \eta \eta  e^{2\phi}b ,\\ \nn
Q_{B}(F)&=&\partial c \xi c G_{m} e^{-\phi},\\ \nn
Q_{0}(V_{1})&=&\fracs{{c_{m}}}{{2}}\fracs{{1}}{{3!}} \partial^{3}c \xi c e^{-\phi}+\fracs{{3}}{{2}}\partial c \xi c T_{m} e^{-\phi}\, ,\\
Q_0(V_{2})&=& 
-\partial \xi c \partial^2 c \ep - \fracs{{3}}{{2}} \xi \de c \de^2 c \ep
- \xi c \de^3c \ep +c\xi \partial^{2}c \partial e^{-\phi},\\ \nn
Q_{0}(V_{3})&=&\fracs{{c_{\xi \eta}}}{{2}}\fracs{{1}}{{3!}}\partial^{3}c \xi c e^{-\phi}+\fracs{{3}}{{2}}c\partial c \xi T_{\xi \eta}  e^{-\phi}+\fracs{{1}}{{2}}c\partial^{2}c \partial \xi e^{-\phi}\,\nn ,\\
Q_{0}(V_{4})&=&\fracs{{9}}{{12}}\partial^{3}c \xi c e^{-\phi}+\fracs{{3}}{{2}}\partial c \xi T_{\phi} c e^{-\phi} + \fracs{{1}}{{2}}\xi \de^2 c c \de \phi \ep\,\nn  ,\\
Q_0 (V_{5})& = & -\fracs{{2}}{{3}} \de^3 c \xi c + \xi c \de^2 c \de \phi \ep + \fracs{{3}}{{2}} \xi c \de c \de^2 \phi \ep ,\\
Q_{0}(U_{A})&=&0\, ,\\
Q_{0}(U_{2})&=&\partial(c \xi \partial c e^{-\phi}) ,\\
Q_0(U_{3})& = & \fracs{{1}}{{2}} \de \xi c \de c \ep \nn ,\\
Q_0(U_{4})& = & \fracs{{1}}{{2}} c \de c \xi \de \phi \ep - \frac{1}{2}
c \de^2 c \xi \ep .\\
\end{eqnarray*}
\begin{center}
\[
c_m=15, \,\,\,\,\,\,\,\,\,\,\,\,\,\,\,\,\,\,\,\,\,\,\,c_{\xi \eta}=-2.
\]
\end{center}
{\bf $Q_{1}$:}
\begin{eqnarray*}
Q_{1}(F)&=&-10\partial \eta \,\xi c -10 \eta \xi c (\partial \phi)-2T_{m}c \partial \phi -5(\partial^{2}\phi +(\partial \phi)^{2})c ,\\ 
Q_{1}(V_{1})&=&\fracs{{3}}{{2}}\xi \eta c G_{m}-\fracs{{3}}{{2}}c G_{m}\partial \phi-\fracs{{1}}{{2}}c\partial G_{m}\,,\\
Q_{1}(V_{2})&=&0 ,\\ \nn
Q_1 (V_{3})&=&\xi \eta c G_m \nn ,\\
Q_{1}(V_{4})&=&-\fracs{{3}}{{2}}G_{m}\eta \xi c -\fracs{{5}}{{2}}G_{m}\partial \phi c -\fracs{{3}}{{2}}\partial G_{m} c ,\\ \nn
Q_{1}(V_{5})&=&G_{m}\eta \xi c +\partial G_{m}c +G_{m}\partial \phi c ,\\ \nn
Q_{1}(U_{2})&=&0,\\ \nn
Q_1(U_{3}) & = & G_m c ,\\ \nn
Q_{1}(U_{4})&=&G_{m}c.\\ \nn
\end{eqnarray*}
{\bf $Q_{2}$:}
\begin{eqnarray*}
Q_{2}(T)&=&-\eta e^{\phi},\\ \nn
Q_{2}(V_{1})&=&-\eta e^{\phi}T_{m},\\ \nn
Q_{2}(V_{2})&=&2\eta \partial \eta \xi e^{\phi}-3\partial^{2}\eta e^{\phi}-8\partial \eta \partial \phi e^{\phi}-2\eta \partial^{2}\phi e^{\phi}-4\eta (\partial \phi)^{2}e^{\phi},\\ \nn
Q_{2}(V_{3})&=&\eta \partial b c e^{\phi}+2\eta b c (\partial \phi)e^{\phi}+3\partial \eta \xi \eta e^{\phi}+\eta \partial^{2}\phi e^{\phi} +2\eta (\partial \phi)^{2}e^{\phi} ,\\ \nn
Q_{2}(V_{4})&=&-4\eta \partial \eta \xi e^{\phi}+8\partial \eta b c e^{\phi} +4\eta \partial b c e^{\phi}+10 \eta b c \partial \phi e^{\phi},\\ \nn
&+&6\partial^{2}\eta e^{\phi}+20 \partial \eta \partial \phi e^{\phi}+5\eta \partial^{2}\phi e^{\phi}+\fracs{{25}}{{2}} \eta(\partial \phi)^{2}e^{\phi},\\ \nn
Q_{2}(V_{5})&=&2\eta \partial \eta \xi e^{\phi}-4\partial \eta b c e^{\phi}-2\eta \partial b c e^{\phi}-4\eta b c \partial\phi e^{\phi}-3\partial^{2}\eta e^{\phi}-8\partial\eta \partial \phi e^{\phi} \nn ,\\  &-&3\eta \partial^{2}\phi e^{\phi}
-4\eta (\partial\phi)^{2}e^{\phi},\\ \nn
Q_{2}(U_{3})&=&-2\eta b c e^{\phi}-3\partial \eta e^{\phi}-4\eta \partial \phi e^{\phi},\\ \nn
Q_{2}(U_{2})&=&-2\partial (\eta e^{\phi}),\\ \nn
Q_{2}(U_{4})&=&-2\eta b c e^{\phi}-5\eta \partial \phi e^{\phi} -4\partial \eta e^{\phi}.\\ \nn
\end{eqnarray*}

\end{document}